\documentclass[%
aps,prl,twocolumn,superscriptaddress
]{revtex4-1}
\usepackage{graphicx}
\usepackage{subfigure}
\usepackage{bm}
\usepackage{hyperref}
\usepackage{indentfirst}
\usepackage{braket}
\usepackage{amsmath}
\hypersetup{colorlinks=true, citecolor=blue, urlcolor=blue, linkcolor=blue}
\pdfoutput=1

\newcommand{\correspondingauthor}{\textsuperscript{$\dagger$}}

\begin{document}
\title{Floquet-Engineered Hybrid Topological Orders with Majorana Edge Modes in Number-Conserving Fermionic Quantum Simulators}

\author{Zhi Lin \correspondingauthor} 
\email[Corresponding author: ]{zhilin18@ahu.edu.cn}
 \thanks{ These authors have contributed equally to this work.}
 \affiliation{School of Physics and Optoelectronic Engineering, Anhui University, Hefei 230601, China}
    
\author{Qi Song }
\thanks{These authors have contributed equally to this work.}
    \affiliation{Department of Physics and State Key Laboratory of Surface Physics, Fudan University, Shanghai 200433, China}
    
\author{Sheng Yue}
\affiliation{School of Physics and Optoelectronic Engineering, Anhui University, Hefei 230601, China}  
\author{Ming Yang}
\affiliation{School of Physics and Optoelectronic Engineering, Anhui University, Hefei 230601, China}  
\affiliation{Institute of Artificial Intelligence, Hefei Comprehensive National Science Center, Hefei 230088, China}
\author{Jie Lou}
\affiliation{Department of Physics and State Key Laboratory of Surface Physics, Fudan University, Shanghai 200433, China}
\author{Yan Chen}
\email[Corresponding author: ]{yanchen99@fudan.edu.cn}
    \affiliation{Department of Physics and State Key Laboratory of Surface Physics, Fudan University, Shanghai 200433, China}

\date{\today}

\begin{abstract}
 We develop an experimental protocol based on Floquet-engineered ultracold fermions in optical lattices, enabling the emulation of pair-hopping and competing singlet/triplet pairing interactions. Through large-scale density matrix renormalization group (DMRG) simulations, we uncover three emergent topological phases: (i) A Majorana-enabled spin-density-wave (MS) phase featuring exponentially localized edge charges, non-local fermionic edge correlations, and doubly degenerate entanglement spectra; (ii) A z-axis polarized triplet superconducting (TS) phase exhibiting fractionalized edge spins (S=1/4 per edge), two-fold ground state degeneracy  and a bulk single-particle gap; (iii) A hybrid x-directional triplet superconducting (XTS) phase that uniquely combines fractional spin textures and Majorana-type edge correlations, defining a new universality class of hybrid orders in number-conserving systems. These findings establish a universal framework for engineering non-Abelian topological matter, crucially bypassing the need for external pairing fields while maintaining experimental feasibility with current cold-atom techniques.
\end{abstract}  

\pacs{}
\maketitle

\emph{Introduction.---} 
During the past decades, the pursuit of topological quantum computing has driven intense investigation of Majorana zero modes (MZMs), whose non-Abelian braiding statistics offer a pathway toward scalable and fault-tolerant qubit operations \cite{1,2,3,4,5,6}. They were first proposed in a one-dimensional system of spinless fermions with $p$-wave superconducting pairing interactions, known as the famous Kitaev model \cite{7}, where particle number is non-conserving. Subsequently, MZMs have inspired numerous implementations in both solid-state heterostructures \cite{9,10} and ultracold atomic platforms \cite{11,12},  typically requiring proximity-induced superconductivity/superfluidity.

Recently, more and more attention has been focused on one-dimensional systems with a number-conserving setting \cite{13,14,15,Wpair2013Zoller,17,18,Wpair2015Buchler,TS2015ErezBerg,21,22,23,24,25,26,27},  as these systems do not rely on an external pairing field. Ref. \cite{15} introduces a minimal particle number-conserving model. In this model, interchain pair tunneling serves as a key factor for MZMs by coupling two chains of a ladder with interacting spinless fermions. Subsequently, this novel pairing interaction was proposed in cold atomic systems by suppressing single-particle hopping by introducing offsets in the optical lattice \cite{Wpair2013Zoller}. However, its experimental realization remains challenging due to excessive laser requirements and requiring adiabatic eliminataion of the auxiliary degrees of freedom.


The bosonic counterpart of this fermionic-type pairing (FTP) interaction, referred to as synthetic pair hopping (SPH) interaction,  was proposed in our previous work \cite{29} through Floquet engineering with a continuous driving protocol in a two-component boson system. Subsequently, a pulse-sequence-based driving protocol \cite{Pulse-sequence-based} has been also theoretically shown to effectively implement the SPH interaction.
Compared to the implementation scheme of the SPH interaction \cite{29}, it is easy to notice that by introducing nearest-neighbor repulsive interactions in a two-component fermion system and using the same driving scheme, we can realize this FTP interaction. In addition, reviewing the process of realizing the SPH interaction reveals that the synthetic anti-pairing interaction ($\hat{a}^{\dag}_{s}\hat{b}_{s}\hat{b}^{\dag}_{s}\hat{a}_{s}$) is inevitably generated simultaneously. Thus, a driving scheme similar to that for the SPH interaction to realize the FTP hopping interaction will inevitably generate simultaneously the fermionic-type anti-pairing hopping interaction. Interestingly, this fermionic-type anti-pairing hopping interaction can be expressed as a combination of spin-singlet pairing and $z$ component of spin-triplet pairing interaction (see a derivation in the Supplementary Material).

On the other hand, searching for new topological phases and novel topological edge states is a fascinating research problem. Our previous analytical study \cite{25} revealed two distinct types of topological phases, i.e.,  $z$-directional triplet pairing superconducting (TS) phase and Majorana-spin-density-wave (MS) phase, in an interacting fermionic two-leg ladder system with FTP interaction, spin-singlet pairing (SSP), and spin-triplet pairing (STP) interactions. Under the mean-filed approximation, the TS phase (MS phase with half filling) is characterized by a fractional edge spin (charge) and can be interpreted as two Kitaev (Su–Schrieffer–Heeger ) chains \cite{25}. A natural question is whether a new topological phase could emerge in this system that simultaneously exhibits the properties of these two states.

In this paper, we propose a Floquet engineering scheme in a one-dimensional spin-fermion system to emulate the FTP, both SSP and STP interactions, aiming to simulate and control novel topologically nontrivial phases in optical lattices. Three topologically nontrivial phases are revealed: the MS phase, the TS phase, and the $x$-directional triplet pairing superconducting (XTS) phase. Intriguingly, the XTS phase displays characteristics of both the MS  and  TS phases. Specifically, the XTS phase exhibits fractional edge spins along the $x$-direction and non-local single-particle edge correlations. Furthermore, various topologically trivial phases are also exposed.

\begin{figure}[t]
\centering
\includegraphics[width=1.0\linewidth]{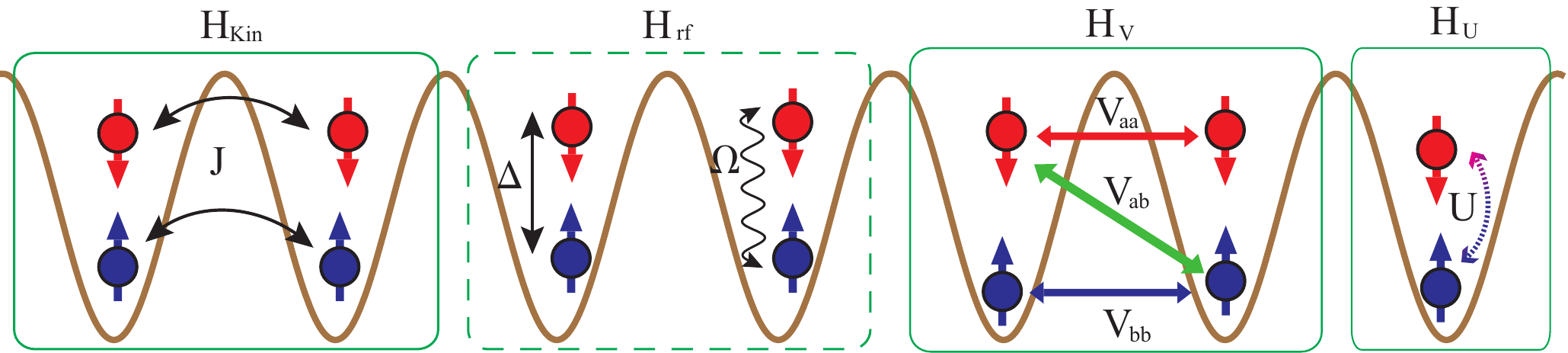}
\caption {The physical processes of the periodic modulated two-component fermion system. Here, $H_{kin}$ denotes the spin-independent hopping between the nearest-neighbor sites, $H_{rf}$ describes the radio-frequency coupling of the two spin states with periodic modulated Rabi frequency $\Omega(t)$, $H_{V}$ is the nearest-neighbor on-site interaction and $H_{U}$ is the on-site interaction. }
\label{schematic}
\end{figure}

\emph{The effective Hamiltonian.---}
We introduce the experimental scheme of the realization of SSP interacting $H_s$, STP interacting $H_{d}$ and pair hopping interaction $W$ for two-component fermions in a one-dimensional (1D) systems (single chain) by using periodic modulating radio-frequency field. Firstly, we present the time-dependent Hamiltonian of this periodic modulated two-component fermion system, and the corresponding physical processes of this periodic modulating system are presented in Fig.~\ref{schematic}. Then the
corresponding time-dependent Hamiltonian can be written as \cite{ZhiLin2017PhDthesis} 
\begin{eqnarray}
H\!\!&=&\!\!-t\sum_{i}(A^{\dagger}_{i}A_{i+1}+H.C.)\!+U \sum_{i} n_{i\uparrow}n_{i\downarrow}\!\nonumber \\
&&+V_{\uparrow\uparrow}\sum_{i}n_{i\uparrow}n_{i+1\uparrow}+V_{\downarrow\downarrow}\sum_{i}n_{i\downarrow}n_{i+1\downarrow}\nonumber \\
&&\!\!+ V_{\uparrow\downarrow}\sum_{i}\left(n_{i\uparrow}n_{i+1\downarrow}+n_{i\downarrow}n_{i+1\uparrow}\right)
\nonumber \\
&&\!\!+\frac{\hbar \Delta}{2} \sum_{i} A^{\dagger}_{i}\sigma_{z}A_{i}-\frac{\hbar \Omega(t)}{2} \sum_{i} A^{\dagger}_{i}\sigma_{x}A_{i},
\end{eqnarray}
where $t$ denotes the spin-independent hopping amplitude, $U$ the on-site Hubbard interaction, and $V_{\downarrow\downarrow}$, $V_{\uparrow\uparrow}$, and $V_{\uparrow\downarrow}$ the nearest-neighbor interactions. $A_{i}^{\dagger}=(c_{i\downarrow}^{\dagger},c_{i\uparrow}^{\dagger})$ is a vector field with creation operators $c_{i\downarrow}^{\dagger}$ ($c_{i\uparrow}^{\dagger}$) on site $i$ of the chain for spin-down (spin-up) component, $n_{i\downarrow}=c^{\dagger}_{i\downarrow}c_{i\downarrow}$ ($n_{i\uparrow}=c^{\dagger}_{i\uparrow}c_{i\uparrow}$) denotes the number operator of the spin-down (spin-up) component, $\Delta=\omega_{A}-\omega_{rf}$ is the detuning of the radio wave ($\omega_{rf}$) with respect to the atomic resonance ($\omega_{A}$), $\Omega(t)=\Omega \sin(\omega t)$ is the Rabi frequency (time-dependent), and $\sigma_{x,y,z}$ denote the Pauli matrices \cite{rf1}. We are interested in the case of $V_{\downarrow\downarrow}=V_{\uparrow\uparrow}=V$ and $V_{\uparrow\downarrow}=V+\delta V$. Therefore nearest-neighbor on-site interaction can be written as $H_{\rm{V}}=H_{\rm{V0}}+H_{\rm{V1}}$, with $H_{\rm{V0}}=V\sum_{i}A^{\dagger}_{i}A_{i}A^{\dagger}_{i+1}A_{i+1}$ and
$H_{\rm{V1}}=\delta V\sum_{i}\left(n_{i\downarrow}n_{i+1\uparrow}+n_{i\uparrow}n_{i+1\downarrow}\right)$.

The effective Hamiltonian under high-frequency approximation can be simplified as (see a
detailed derivation in the Supplementary Material \cite{SM})
\begin{eqnarray}
\mathcal{H}_{\rm{eff}}\!\!
&=&\!\!-t\sum_{i}(A^{\dagger}_{i}A_{i+1}\!+\!H.C.)\!\nonumber \\
&&\!\!+W\sum_{i}(c^{\dagger}_{i\uparrow}c^{\dagger}_{i+1\uparrow}c_{i\downarrow}c_{i+1\downarrow} + c^{\dagger}_{i\downarrow}c^{\dagger}_{i+1\downarrow}c_{i\uparrow}c_{i+1\uparrow}) \nonumber \\
&&\!\!+g_{s}\sum_{i}\!\Delta^{\dagger}_{si}\Delta_{si}+g_{d}\sum_{i}\!\Delta^{\dagger}_{di}\Delta_{di}+U\!\sum_{i}\! n_{i\uparrow}n_{i\downarrow}\!\!\nonumber \\
&&\!\!+V_{\rm{eff}}\!\sum_{i}\!\left(n_{i\downarrow}\!+\!n_{i\uparrow}\right)\!(n_{i+1\downarrow}\!+\!n_{i+1\uparrow}),
\label{eff-Hamil}
\end{eqnarray}
where the preceding four terms have been studied in our previous work \cite{Liu2019}, the last two terms are the on-site interaction and nearest-neighbor on-site interaction. Here we have $W=-\delta V\left(\frac{\Omega}{2\omega}\right)^{2}$ and $g_{s}=\frac{\delta V}{2}$, $g_d=\frac{\delta V}{2}[1-4\left(\!\frac{\Omega}{2\omega}\right)^{2}]$, $V_{\rm{eff}}=\left(V+\delta V\left(\!\frac{\Omega}{2\omega}\!\right)^{2}\right)$, $\Delta_{s(d)i}=c_{i\uparrow}c_{i+1\downarrow}\mp c_{i\downarrow}c_{i+1\uparrow}$. In practical calculations, for ease of study, we set $V_{\rm{eff}}=U=0$. These conditions can be realized in current experiments using Feshbach resonance techniques \cite{Feshbach1, Feshbach2,schafer2020tools}.


\begin{figure}
\centering
\includegraphics[scale=0.6,trim=0 0 0 0,clip]{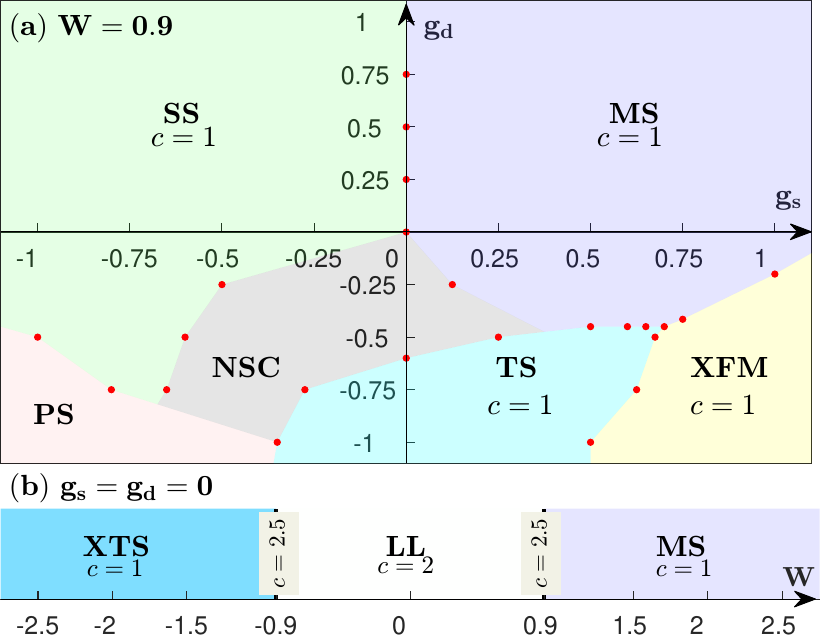}
\caption{\label{fig_PD}
Approximate phase diagram for the $L=240$ system at density $\rho=0.4$. 
(a) For positive $W=0.9$, as $g_s$ and $g_d$ vary, six phases appear: Majorana-spin-density-wave (MS) phase, triplet pairing superconducting (TS) phase, $x$-directional ferromagnetic (XFM) phase, singlet pairing superconducting (SS) phase, normal superconducting (NSC) phase with coexisting triplet and singlet pairing, and phase separation (PS). 
Different colors represent the various phases, and the red circles indicate the transition points corresponding to the central charge $c$ values.
(b) For $g_s=g_d=0$, there are three phases: $x$-directional triplet pairing superconducting (XTS) phase, Luttinger Liquid (LL) phase, and MS phase. The transition points at $W=\pm 0.9$ exhibit an enhanced central charge $c=2.5$. 
The TS (cyan area), MS (purple area), and XTS (dark blue area) phases are topological, whereas the other phases are topologically trivial. 
}
\end{figure}

\emph{Phase diagram.---}
Before studying the phase diagram of this strongly interacting system, let us first briefly review its behavior under weak coupling. The ground-state phase diagram in the weak coupling limit has been studied in our previous work by the Abelian bosonization method \cite{Liu2019}. The bosonization analysis predicts two topologically nontrivial phases, the TS and MS phases, and a topologically trivial phase with singlet pairing superconducting (SS) order. Negative $g_d\ (g_s)$ favors the singlet (triplet) pairing superconducting phase. The TS phase is characterized by a fractional edge spin $S^z=\pm 1/4$. The pair hopping supports spin-density-wave orders and leads to Majorana edge modes. The resulting MS phase supports exponentially localized edge charge and exhibits non-local correlations between the two ends of the system.
To study the ground-state phase diagram under strong coupling, we employ the state-of-the-art density matrix renormalization group (DMRG) method with conserved total particle number $N=N_{\uparrow}+N_{\downarrow}$ \cite{PhysRevLett.69.2863, ITensor}. 
The system size is up to $L=240$ with open boundary conditions (OBC).
We choose $t=1$ as the unit of energy and the total particle density $\rho=N/L=0.4$ with $N_{\uparrow}=N_{\downarrow}$, if not specified. 

Apart from the above three phases, DMRG calculations give a richer phase diagram shown in Fig. \ref{fig_PD}. For $W<0$, we find a novel topological XTS phase that features fractional edge spin along $x$-direction and non-local single-particle edge correlations. The topological XTS, TS, and MS phases are protected by time-reversal symmetry (TRS) and spin parity symmetry ($P_s$) \cite{Liu2019}, each exhibiting twofold degenerate ground states under OBC.
For $W>0$, we identify a normal superconducting (NSC) phase with coexisting triplet and singlet pairing, as well as a ferromagnetic superconducting (XFM) phase with spin polarization along the $x$-direction. Further details of the NSC, XFM, and SS phases are provided in the Supplementary Material \cite{SM}. 

The phase diagram for $W>0$ is depicted in Fig. \ref{fig_PD}(a), according to the changes of the central charge $c$. In ($1+1$)-dimensional conformal field theory, the central charge can be extracted from the fits of the entanglement entropy $S_{vN}$, defined as $S_{vN}=-$Tr [$\rho_l$ ln$ \rho_l]$ where $\rho_l$ is the reduced density matrix of a subsystem of length $l$ \cite{Calabrese_2009, Calabrese_2004, SM}. For $g_d<0$ and small $g_s$, the system in the TS phase exhibits a single gapless bosonic mode due to a strong triplet superconducting order parameter (corresponding to $c=1$), while its spin sector is gapped. Such a gapless mode resembles gapless topological superconductors proposed in previous studies with spin-orbital coupling \cite{TS2015ErezBerg}. As $g_d$ increases for positive $g_s$ in Fig. \ref{fig_PD}(a), the system undergoes a phase transition from the TS phase to the MS phase ($c=1$) with a central charge $c=2$ transition point. $c=2$ relates to two gapless modes, both in charge and spin channels. A finite gap to single fermion excitations in the bulk protects the localized edge states in the TS and MS phases. At the transition point, the single-particle charge gap vanishes, manifested by quasi-long-range single-particle correlation functions, and the topological characteristics also disappear. 

When $g_d<0,g_s>0$ and either one of their absolute values is comparable to $|W|$, the spin-polarized XFM phase appears with $c=1$ as in Fig. \ref{fig_PD}(a). The XFM phase transmutes into the TS or MS phase with a $c=2$ transition point. Furthermore, when the negative $g_s$ prevails, the ground state is in the SS phase with $c=1$. Between the SS and TS phase lies the NSC phase (see Fig. \ref{fig_PD}(a)), where the values of central charge range from 1 to 2.5 as $g_s$ and $g_d$ vary. The topologically nontrivial TS and trivial NSC phases are separated by a phase transition point with an enhanced central charge close to 2.5. This implies the emergence of a gapless Ising or Majorana mode with $c=1/2$ in addition to two standard gapless modes \cite{TwoFluid2021,cc32spinless2017}. 
Additionally, the enhancement of central charge $c=2.5$ is also observed at the transitions between the SS and MS phases, as well as between the Luttinger Liquid (LL) \cite{tJMoreno2011, Jiang2022tripletSC,giamarchi2003quantum,haldane1981luttinger} and XTS phases\cite{SM}. In the following, we will characterize the properties of three distinct topological states in the phase diagram.

\begin{figure}
\centering
\includegraphics[scale=0.55,trim=0 0 0 0,clip]{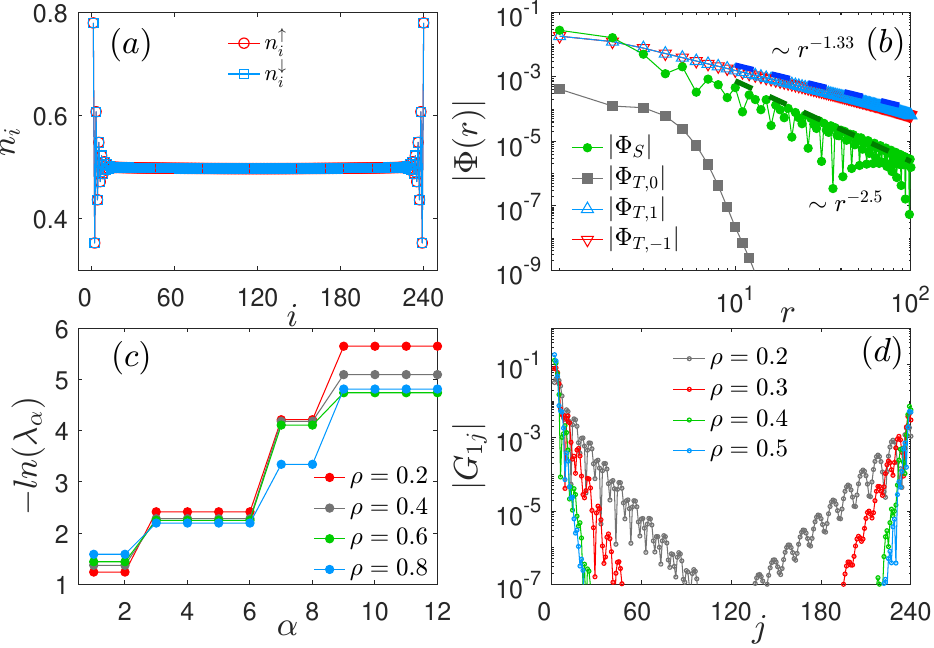}
\caption{\label{fig_MS} 
Characterization of the topological MS phase at $W=0.9,g_s=0.75,g_d=0.25$ for $L=240$. (a) Charge density profile $n_i$ at site $i$ for density $\rho=1.0$. (b) Superconducting pairing correlations $\Phi(r)$ on double-logarithmic scales for density $\rho=0.4$. $r$ is the distance from the reference site $L/4$. The dashed lines denote power-law fitting to $|\Phi(r)|\sim r^{-K_{SC}}$. (c) Degenerate entanglement spectrum for different densities. (d) Non-local single-particle correlations $G_{1j}$ for spin-$\downarrow$ particles between the edge site $1$ and site $j$ with semilogarithmic scale at different densities. $G_{1j}$ for spin-$\uparrow$ particles show the same behavior.
}
\end{figure}

\begin{figure}
\centering
\includegraphics[scale=0.55,trim=0 0 0 0,clip]{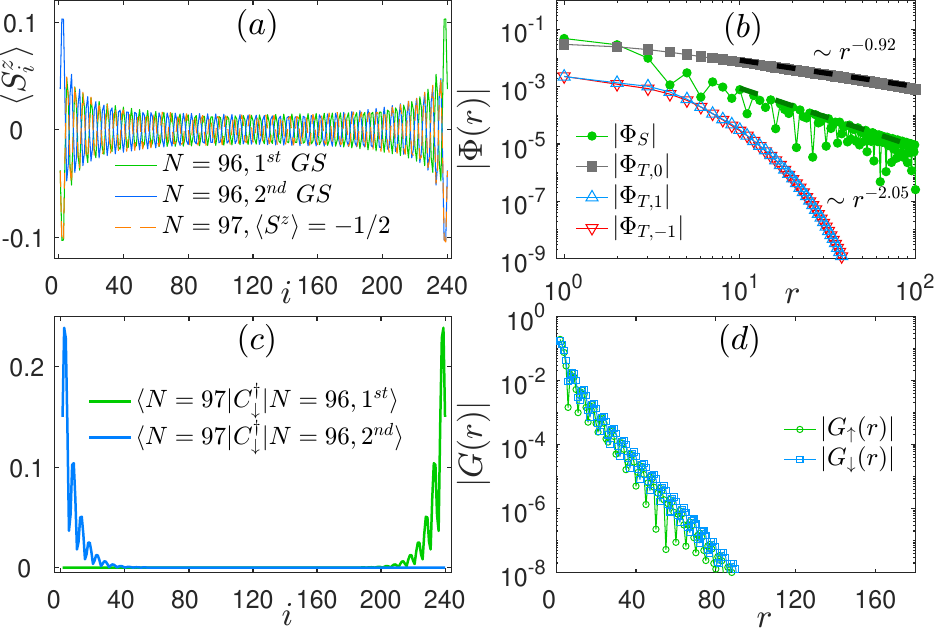}
\caption{\label{fig_TS} Characterization of the topological TS phase at $W=0.9,\ g_s=0.225,\ g_d=-0.675$ for $L=240,\ \rho=0.4$.
(a) $z$-component spin $\langle S^z_i\rangle$ at site $i$. The blue and green solid lines depict the two degenerate GSs of a system with $N=96$ particles and zero net spin. At the system edges, spin accumulation is observed with the integrated spin in the left and right halves being $+0.2348$ ($-0.2348$) and $-0.2348$ ($+0.2348$) respectively for the $1^{st}$GS ($2^{nd}$GS).
The orange dashed line depicts one of the GSs of the $N=97$ system, with a net spin of $ \langle S^z\rangle=-1/2$ and an integrated spin of $1/4$ in each half. 
(b) Superconducting pairing correlations $\Phi(r)$ on double-logarithmic scales. The dashed lines denote power-law fitting $|\Phi(r)|\sim r^{-K_{SC}}$.
(c) Existence of low-energy single-particle states at the edges. The matrix elements represent the overlapping between the two GSs with $N=96$ particles and the GS with $N=97$ particles by adding a spin-$\downarrow$ particle at site $i$.
(d) Exponentially decaying single-particle correlations $G_{\sigma}(r)$ in the bulk on the semilogarithmic scale. 
}
\end{figure}

\emph{Topological MS phase.---}
To characterize distinct quantum phases, we measure charge densities, entanglement spectrum\cite{laflorencie2016quantum, ES2010PollmannBergOshikawa}, spin-spin correlations, single-particle and superconducting pairing correlations. As shown in Fig. \ref{fig_MS}(a), edge charge accumulation in the MS phase follows an exponential density profile. The topological nature of the MS phase also manifests itself in the doubly degenerate entanglement spectrum (ES) and non-local single-fermion correlations $G^{\sigma}_{1j}$ between the edges. The eigenvalues $\lambda_{\alpha}$ ($\alpha=1,...,\chi$) of the reduced density matrix $\rho_{L/2}$ with dimension $\chi$ yield the ES as $-ln\lambda_{\alpha}$. In Fig. \ref{fig_MS}(d), $G^{\sigma}_{1j}=\langle c^{\dagger}_{1\sigma}c_{j\sigma} \rangle$ ($j\in [1, L],\sigma=\uparrow,\downarrow$) attains considerable values around marginal sites while decays exponentially into the bulk, indicating localized Majorana edge states. Additionally, the ground state doubly degenerates due to selecting different local fermionic parities \cite{Wpair2015Sebastian,Wpair2015Buchler}.

Unlike higher-dimensional systems with long-range order, 1D systems exhibit quasi-long-range order (QLRO), characterized by power-law decaying order parameter correlations. The various superconducting pairing correlations are defined as follows: the spin-singlet pairing $\Phi_S(r)=\langle \Delta^{\dagger}_S(i)\Delta_S(j) \rangle$ with $\Delta_S(j)=\frac{1}{\sqrt{2}}(c_{j\uparrow}c_{j+1\downarrow}-c_{j\downarrow}c_{j+1\uparrow})$, and the triplet pairing $\Phi_{T,s}(r)=\langle \Delta^{\dagger}_{T,s}(i)\Delta_{T,s}(j) \rangle$ with $ \Delta_{T,1}(j)=c_{j\uparrow}c_{j+1\uparrow}$, $\Delta_{T,0}(j)=\frac{1}{\sqrt{2}}(c_{j\uparrow}c_{j+1\downarrow}+c_{j\downarrow}c_{j+1\uparrow})$, $ \Delta_{T,-1}(j)=c_{j\downarrow}c_{j+1\downarrow}$, where $r=|j-i|$ is the distance between site $i$ and site $j$. In the MS phase, $\Phi_{T,\pm 1}(r)$ decay in power law $\sim r^{-K_{SC}}$ with the superconducting exponent $K_{SC}\simeq 1.33$ in Fig. \ref{fig_MS}(b). $K_{SC}<2$ implies a divergent superconducting susceptibility at zero-temperature limit \cite{Kivelson2004,Jiang2022tripletSC}. Furthermore, the single-particle correlations are suppressed in the bulk, showing exponential decay. This indicates a nonzero gap for bulk single-fermion excitations, which guarantees the stability of the zero-energy edge modes. Fourier transforms of the spin-spin correlations $F(r)=\langle\boldsymbol{S}_i \cdot \boldsymbol{S}_j\rangle$ show a $2k_F$ spin-density wave (SDW) with Fermi momentum $k_F=\pi\rho/2$ \cite{SM}. 
This feature is distinct from the previously discussed topologically $p$-wave superconducting phase induced by pair hopping\cite{,Wpair2015Buchler, Wpair2013Zoller,Wpair2015Sebastian}.

\emph{Topological TS phase.---}
The predominant QLRO in the TS phase is characterized by zero spin-$z$ triplet pairing superconducting correlations $\Phi_{T,0}(r)$, with a power exponent $K_{SC}\simeq 0.92$, as depicted in Fig. \ref{fig_TS}. This phase coexists with a weak singlet superconductivity, exhibiting a $K_{SC}$ slightly larger than 2. 
The single-particle correlations $G_{\sigma}(r)=\langle c^{\dagger}_{i\sigma}c_{j\sigma} \rangle$ show exponential decay behavior due to the absence of gapless fermionic mode in the bulk. At the system edges, the accumulation of $z$-component spin is observed with an opposite sign between the two degenerate ground states (GSs). Because there is an anomalous relation between time-reversal and local fermion parity operator \cite{TS2015ErezBerg}. If the total number of fermions $N$ is even, total $\langle S^z \rangle = 0$ and the integrated spin in each half of the system is close to half of an electron spin. If $N$ is odd, the two degenerate GSs with total $\langle S^z \rangle = \pm 1/2$ host an exact integrated spin $\langle S^z_{half} \rangle = \pm 1/4$ in each half.  As illustrated in Fig. \ref{fig_TS}(a), the aforementioned edge spin polarization can induce SDW in bulk. 
To further substantiate the presence of a low-energy edge state, we add spin-$\downarrow$ fermion at site $i$ into each of the two GSs with an even particle number $N$. Subsequently, we compute the overlap with one of the GSs with a total particle number of $N+1$.
The value of the overlap is nonzero at only one end depending on which GS is involved, clearly verifying a low-energy single-particle state localized at the edge (see Fig. \ref{fig_TS}(c)).

\begin{figure}
\centering
\includegraphics[scale=0.5,trim=0 0 0 0,clip]{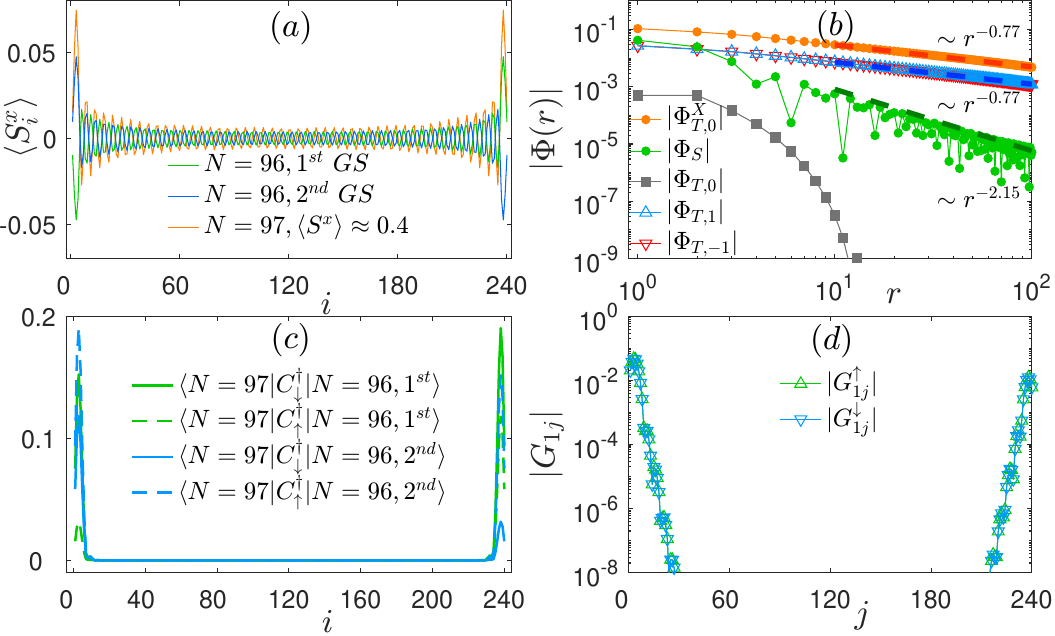}
\caption{\label{fig_XTS}  Characterization of the topological XTS phase at $W=-2.0,\ g_s=g_d=0$ for $L=240,\ \rho=0.4$.
(a) $x$-component spin $\langle S^x_i\rangle$ at site $i$. The blue and green solid lines depict the two GSs of a system with $N=96$ particles and zero net spin. 
At the system edges, spin accumulation is observed with the integrated spin in the left and right halves being $+0.1155$ ($-0.1155$) and $-0.1155$ ($+0.1155$) respectively for the $1^{st}$ GS ($2^{nd}$ GS).
The orange solid line depicts one of the GS of the $N=97$ system, with a net spin of $ \langle S^x\rangle=0.3675$ and an integrated spin of $\langle S^x\rangle/2$ in each half. 
(b) Superconducting pairing correlations $\Phi(r)$ in the bulk on double-logarithmic scales. The dashed lines denote power-law fitting $|\Phi(r)|\sim r^{-K_{SC}}$. $\Phi^X_{T,0}$ is triplet pair-field correlation in the $x$-direction. 
(c) Existence of low-energy single-particle states at the edges. The matrix elements represent the overlapping between the two GSs with $N=96$ particles and the GS with $N=97$ particles by adding a spin-$\uparrow$ or spin-$\downarrow$ particle at site $i$.
(d) Non-local single-particle correlations $G_{1j}$ between the edge site $1$ and site $j$ on semilogarithmic scale.
}
\end{figure}

\emph{Topological XTS phase for $W<0$.---}
When the amplitude of pair hopping $W$ is negative and far away from zero, another kind of quasi-long-range triplet pairing order in the $x$-direction ($\Phi^X_{T,0}$) is observed, in addition to the aforementioned $S^z=\pm 1$ triplet pairing superconducting order ($\Phi_{T,\pm 1}$) (see Fig.\ref{fig_XTS}(b)). The $x$-directional triplet pairing correlation function is defined as $\Phi^X_{T,0}(r)=\langle \Delta^{X\ \dagger}_{T,0}(i)\Delta^X_{T,0}(j) \rangle$ with $\Delta^X_{T,0}(j)=\frac{1}{\sqrt{2}}(c^X_{j\uparrow}c^X_{j+1\downarrow}+c^X_{j\downarrow}c^X_{j+1\uparrow})$ where $r=|j-i|$. Then we can reformulate the $S^x=0$ triplet pair-field as $\Delta^{X}_{T,0}(j)=\frac{1}{\sqrt{2}}(c_{j\uparrow}c_{j+1\uparrow}-c_{j\downarrow}c_{j+1\downarrow})$ by transforming the fermionic operator $c^X_{\sigma}$ in the spin-$x$ basis into a representation in the spin-$z$ basis : $c^X_{\uparrow}=\frac{1}{\sqrt{2}}(c_{\uparrow}-c_{\downarrow}), c^X_{\downarrow}=\frac{1}{\sqrt{2}}(c_{\uparrow}+c_{\downarrow})$. Additionally, the $\Phi^X_{T,0}$ exhibits power-law decay behavior in the XTS phase, while it decays exponentially in the phases appearing for positive $W$. These three types of $p$-wave pairing give rise to unprecedented unique topological properties, including fractional edge spin in the $x$-direction, simultaneous low-energy single-particle excitations at two edges, non-local single-particle edge correlations, and finite single-particle excitation gap in the bulk.    

In Fig. \ref{fig_XTS}(a), the spin accumulation of the $x$-component at edge sites is displayed. The sign of the spin $\langle S^x_i \rangle$ at each site is opposite between the two degenerate GSs, denoted as $1^{st}$GS and $2^{nd}$GS. The integrated spin $\langle S^x_{half} \rangle$ in each half of the system depends on the system parameters. For instance, in a system with even $N=96$ and total $\langle S^x \rangle=0$, $\langle S^x_{half} \rangle=\pm 0.1155$ at $W=-2$, while at $W=-2.5$, $\langle S^x_{half}\rangle$ increases to $\pm 0.2469$, which is nearly half of an electron spin. In a system with odd $N=97$, the two degenerate GSs exhibit a total $\langle S^x \rangle\approx \pm 0.4$ at $W=-2$, while at $W=-2.5$, the total $\langle S^x \rangle$ approaches $\pm 1/2$ \cite{SM}. 

Furthermore, when adding an extra spin-$\sigma$ fermion at site $i$ into each of the two GSs with an even particle number $N$, the overlap with a system with total number $N+1$ and positive total $\langle S^x \rangle$ has a nonzero value at both two edges. At the same time, it becomes zero in the bulk. This concretely confirms the existence of a low-energy state at two edges and further affirms the non-local correlations between the edges (see $G^{\sigma}_{1j}$ in Fig. \ref{fig_XTS}(d)). The value of each overlap is not equal for the two edge sites and spin-$\sigma$, as shown in Fig. \ref{fig_XTS}(c), and this can be used to distinguish between the two degenerate GSs. For example, the overlap involving the $1^{st} GS$ by adding a spin-$\downarrow$ particle shows a more significant portion at the right end. In contrast, the overlap involving the $2^{nd} GS$ by adding up a spin-$\uparrow$ particle shows a larger portion at the left end.

\emph{Experimental realization and detection.---} 
By selecting $\left[\Omega/(2\omega)\right]^{2}\!\!=\!\!0.05\!\ll\!1$, the FTP interaction and fermionic-type anti-pairing become dominant, while the higher-order terms ($\mathcal{O}\!\left[\!f^{4}(t)/\hbar^{4}\right]$) can be neglected \cite{SM}. In this case, the Hamiltonian in Eq.~(\ref{eff-Hamil}) accurately describes all relevant physical processes in this driven system. Recently, the non-local correlations for Majorana edge states can be detected via time-of-flight (TOF) shadow images, where this non-local correlation will induce fast oscillation in TOF images \cite{Zoller2012probeMZM,nascimbene2013probeMZM}. Moreover, the fractional edge charges and fractional edge spins can also be measured \cite{Braun2024, Yao2024, Bloch2022edgeSpinHaldane} using quantum gas microscopy techniques, which provide single-site resolution to map edge spin textures and charge densities directly \cite{microscope1,schafer2020tools,microscope2}. The characteristics of our revealed three topologically non-trivial states are expected to be observed with current experimental techniques.

\emph{Conclusions.---}
We have designed a periodically driven experimental scheme that can simulate the MS phase featuring  MZMs in particle-number-conserving systems and simulate two other types of topological states: the TS phase and the XTS phase. Particularly intriguing is the XTS phase, which exhibits characteristics of both the MS and TS phases, namely fractional edge spins along the $x$-direction and non-local single-particle edge correlations. Furthermore, the properties of these topologically non-trivial states can be measured with current experimental techniques. Through numerical calculations using DMRG, we have provided the phase diagram of the system under several parameters and determined the central charge of various gapless states, as well as the central charge at the boundaries between the XTS and LL phase, and between the LL and MS phases. The central charge at these boundaries is 2.5, indicating the presence of gapless edge excitations. Our proposal enables the simulation of these three novel types of topological states and various topologically trivial states. 

\vspace{0.2cm}
\emph{Note added.---}
During the revision process, we noted the preprint~\cite{Goldman2025}, which presents a similar driving scheme for generating pair-hopping processes in a fermionic ladder system. Their numerical studies reveal a topological ground state featuring MZMs, reminiscent of our MS phase but lacking spin-density-wave order. Furthermore,  a novel XTS phase has been discovered in our work, which uniquely supports both fractional edge spins and MZMs.


\begin{acknowledgements}
\emph{Acknowledgements.---}
     This work is supported by the National Key Research and Development Program of China Grant No. 2022YFA1402204 and the National Natural Science Foundation of China Grant Nos. 12274086, 12004005, the Scientific Research Fund for Distinguished Young Scholars of the Education Department of Anhui Province No.2022AH020008, the Natural Science Foundation of Anhui Province under Grant Nos. 2008085QA26 and 2008085MA16, the University Synergy Innovation Program of Anhui Province under Grant No. GXXT-2022-039, and the open project of the state key laboratory of surface physics in Fudan University under Grant No. KF2021$\_$08.

\end{acknowledgements}

\bibliography{topo}

\end{document}